\documentclass[a4paper]{article}

\usepackage[english]{babel}
\usepackage[utf8x]{inputenc}
\usepackage[T1]{fontenc}

\usepackage[a4paper,top=3cm,bottom=2cm,left=3cm,right=3cm,marginparwidth=1.75cm]{geometry}

\usepackage{amsmath}
\usepackage{graphicx}
\usepackage[colorinlistoftodos]{todonotes}
\usepackage[colorlinks=true, allcolors=blue]{hyperref}

\title{Towards Methods for Model-Based Software Development}
\author{Valdemar Vicente Graciano Neto and
Luiz Fernando Batista Loja\thanks{Instituto de Inform\'atica (INF/UFG), Universidade Federal de Goi\'as, Goi\^ania, Brazil; ICMC, University of S\~ao Paulo (USP), S\~ao Carlos, Brazil;Instituto Federal de Goi\'as (IFG), Luzi\^ania, Brazil. Email: valdemarneto@inf.ufg.br, luizloja@gmail.com}
}

\begin{document}
\maketitle

\begin{abstract}
Software engineering is a young discipline. Despite efforts in recent decades, some elements still require further development, research, and systematization. One of these elements are methods for software engineering. They consist of a set of well-defined activities used in a software development to guide how the work should be performed to achieve the expected results. However, there is a lack of systematic knowledge that effectively guide how work should be done in a variety of areas. Despite the knowledge available in Software Engineering Body of Knowledge (SWEBOK), only the classical methods are comprised there, while emerging topics such as Agent-Oriented Software Engineering (AOSE), Aspect-Oriented Software Development (AOSD), and Model-Based Software Development (MBSE) are only briefly mentioned. We claim that it is also necessary to investigate methods for those emerging software engineering subtracks. In this direction, this paper presents a a speculative and preliminary emerging results on the establishment of methods for MBSE. We report some insights on the conception of methods for MBSE.
\end{abstract}

\section{Introduction}

Software Engineering is a young engineering. Despite the success achieved in the last decades, the abstract nature of software has been challenging for predictions of cost, time, quality. There is also a lack of systematization on procedures, processes, and methods to guide a successful software development. Software Engineering Body of Knowledge (SWEBOK) is one among other attempts to systematize the Software Engineering's knowledge \cite{pontus:2012, swebok:2014}. Most of the Software Engineering subareas are covered there. However, emerging paradigms are not covered, yet. Feature-Driven Development (FDD) and Model-Based Software Development (MBSE) are only briefly cited, while Agent-Oriented Software Engineering (AOSE) is not mentioned \cite{swebok:2014}. Software  engineering  methods  provide  an  organized and systematic approach to developing software. There are numerous methods  to choose, and it is important  for a software engineer to choose an appropriate one or a set of them for the development at hand. Nevertheless, such choice has a dramatic effect on the success of the software project \cite{swebok:2014}. 

Despite the importance of such topic, there is a lack of details on how they should be structured, which recurrent activities comprise them, how they are related to the software development processes, and how they could be associated to produce a better result. Thus, in those situations, software engineers come back again to the pragmatism and experience-based decisions, since there is no a theoretical basis to ground their decisions. SEMAT (Software Engineering Method and Theory)\footnote{\url{http://semat.org/}} initiative is a remarkable instance of means to systematize methods and knowledge in software engineering. However, for recent areas such as MBSE, there is still a lack of systematization on methods. 

We have invested efforts on MBSE in the last years  \cite{bwMDD2010iu,bwMDD2010bpm,earlyaspects:2011,bwMDD2011, sbsi:2013, sbsi:2014en, laweb:2014, GracianoNeto2014b, GracianoNeto:2015, GracianoNeto2016SRCPosBlind, GracianoNeto2017, GracianoNeto:HICSS2018}. Despite MBSE successful cases \cite{DiRuscio:2014, Hutchinson:2014, buttner:2014, Cuadrado:2014}, there is still a lack of consensual set of techniques to develop a metamodel, for example, or to engineer a bidirectional transformation. Indeed, in software engineering, one of the most hotly debated questions concerns the choice of method \cite{pontus:2012}. And, when you reduce this scope for MBSE, we claim that the same occurs. 

This paper presents preliminary insights on methods for MBSE. We present a brief discussion on the basis which could be followed to edify a Methods Theory for MBSE. Aiming at suitably present the content, this paper is structured as follows: Section II presents some foundations to foster and feed the discussion, Section III presents our initial perceptions on the topic, and Section IV presents a brief discussion, while Section V brings final considerations and discusses future work perspectives.

\section{Background}


Essentially, a software development process produces models that are successively transformed along the process until reaching software code. In this scenario, Model-Based Software Development (MBSE) emerged as a solution where models are the first-class citizen \cite{sendall:2003, Hutchinson:2011, Hutchinson:2011b, amrani:2012}. Model transformations are performed in an automated way, through the use of specific software tools \cite{bosems:2008, STVR:2014}. Now, the program is the model \cite{cuadrado:2013}
Adopting MBSE requires artifacts and tools. In MBSE, remarkable artifacts involve models, metamodels, model transformers, and model transformations. A \textbf{model} is, essentially, an abstraction of reality represented in a textual or visual language (usually, a Domain-Specific Language (DSL)). \textbf{Metamodels} are special models that guides and restricts how to construct other models. Models necessarily need to conform to their respective metamodels \cite{conformsto, Cicchetti:2008}. \textbf{Model transformers} are considered model compilers which receive models and their respective metamodels as input and, through the use of \textbf{model transformations}, they transform source models in target models. 	MBSE can be considered one of the innovative software development processes recently created. As such, MBSE is also based on processes, methods, activities, and technologies which support them.


In parallel, processes, methods, techniques, and activities are part of Software Engineering. A process is a framework which involves methods, techniques, activities, and tools \cite{swebok:2014}. A process can have many methods associated (i.e., a process can have Object-Oriented method and Agile methods applied in parallel). A method can be a part of zero or many processes (i.e., formal methods are used in some processes, but many software development processes do not use it). One method has at least one inherent technique (to exist), or many techniques related to it (i.e., classes identification via grammar classes for OO-method, stand-up meetings for agile methods, formal specification for formal methods). And one technique can be part of many processes, as domain engineering, that analyses commonalities and variabilities, and can be a part of feature-driven development, or product-line software development.

Methods should be enough to predict in the each stage of software developments the quality of the end product and respective techniques which should be adopted to achieve success and to avoid repetitive labor and error \cite{pontus:2012}. However, this is not true yet for software engineering, specially for MBSE. In SWEBOK, methods theory is structured in four well-delimited categories:  Heuristic Methods (Structured paradigm, data modeling paradigm, and object-oriented paradigm), Formal Methods (program specification and validation, and program refinement and derivation), Prototyping Methods (styles, target, and evaluation techniques), and Agile Methods (RAD, XP, Scrum, and FDD) \cite{swebok:2014}. Significant progress has been made over the past decades in establishing methods and techniques that support the demands on software development processes. However, there is still a need for improving those methods and techniques \cite{weiss:2009}, besides proposing new ones for emerging paradigms such as MBSE.


%






\section{Methods for MBSE}

For have observed pragmatic issues and acquired experience on MBSE in the last years. With such experience, we tried to gather part of this to structure and  potentially generalize our knowledge \cite{smolander:2013}. A first step to characterize methods is exposing the inherent techniques and activities that recurrently compose them. When considering techniques, they are usually embodied in a method, and they are supported by tools \cite{swebok:2014}. From the best of our knowledge, there is a lack of studies that characterize MBSE techniques, and group them in methods. Hence, we struggle to elicit a set of techniques which characterizes a \textit{Model-Driven Method (MDM)} for each potential stage of a software development process. Then, we can catalog a set of techniques to characterize a MDM for both flows: to construct MBSE tools, and another to use MBSE tools.

The main activity of MBSE is \textit{modeling}. Consequently, every technique related to modeling activity could be added to a MDM. Another important activity in MBSE is \textit{metamodeling}. Similarly, every technique related to metamodeling activity could be added to a MBSE method. Then, we could list at least these main activities in MBSE: modeling, metamodeling \cite{Cuadrado:2012}, checking consistency between model and metamodel \cite{GuerraLWKKRSS13}, transforming models to models, and transforming models to code. 

Tables \ref{tab:process1} and \ref{tab:process2} summarize what could be an MDM: Table \ref{tab:process1} shows the MBSE tools engineering flow, its activities, and each technique that characterize the MDM for that activity. Table \ref{tab:process2} shows the techniques that characterizes MDM use in a classic software development process for each activity of a development process.

Modeling in this context is the activity of \textit{instantiating} a metamodel to obtain a concrete model. And metamodeling is the activity of performing a domain engineering to elicit and document all the concepts that characterizes an specific domain that should be included in a conceptual model for that domain, in such a way that concrete models could be \textit{instantiated} from them, keeping the conformance relation.

\begin{table}

	\centering
		\begin{tabular}{|p{5cm}|p{6cm}|}
		  \hline \textbf{Activity} & \textbf{Techniques} \\\hline
			 Metamodel Specification & Domain Engineering  \\ \hline
			 Metamodel Design & Separation of Concerns in specific models, Concepts representation   \\ \hline
			 Metamodel Implementation & Converting in a textual language (tangled with a transformer implementation, or modularized in a XML, or other structure)  \\\hline 
			 Model Transformation Engineering & Linking one or more source metamodel concepts to one or more target metamodel concepts. \\\hline
			\end{tabular}
	\caption{MBSE engineering flow and its techniques.}
	\label{tab:process1}
\end{table}

\begin{table}
	\centering
		\begin{tabular}{|p{5cm}|p{6cm}|}
		  \hline \textbf{Activity} & \textbf{Techniques} \\\hline
			 Requirements Engineering & Modeling (instantiating a metamodel), checking consistency between model (with tool support or not)  \\ \hline
			 Design & Transform models to models   \\ \hline
			 Codification & Transform models to code  \\\hline 
			 Testing & Using models to extract test cases and test them \\\hline
			\end{tabular}
	\caption{Application Engineering and MDM techniques}
	\label{tab:process2}
\end{table}


It is important to highlight that tables present only techniques that could be considered inherent to MDM application. Analogously to mentioned before, other techniques from other methods as Object-Oriented, Agile, for Formal could be used in association with MDM application.

\section{Discussion}

MBSE has not enough maturity about a taxonomy of methods and techniques, and efforts has been performed to establish MBSE as a complete software development process, with a set of methods and their respective set of techniques.  SWEBOK \cite{swebok:2014} defines a method as an organized and systematic approach for developing software for a target computer. According to SWEBOK, methods can be classified as Heuristic Methods (Procedural, Object-Oriented, and Data Modeling), Formal Methods, Prototyping Methods, and Agile Methods. From the best of our knowledge, there is no study in the literature that characterizes MBSE as a method in the strict meaning of the word in Software Engineering under SWEBOK perspective. 

Considering that classification, MBSE can be considered itself as new method that is not mentioned yet because did not reach enough maturity (in a set that includes Agent-Oriented Software Development, Aspect-Oriented Software Development, Feature-Oriented Software Development, and any another new technology that originates new software development paradigms). On the other hand, there some occurrences of associations between these main method areas and MBSE, as Agile-MBSE \cite{ambler:2003, basso:2010}, and Formal MBSE \cite{Oquendo:2006, oquendo07}.  Additionally, there is no study that consider MBSE as a software engineering method. MBSE is closer to the context of software development process \cite{sbsi:2013}. It involves processes, methods, and techniques, as other software development process. However, we understand that most of the software development life cycle has a method associated: Agile and agile method, and prototyping and prototyping method. In fact, there is still a need for a consensual vocabulary to describe issues such as paradigm, method, process, technique, approach, methodology, since those words often are used interchangeably.

Mature  academic  disciplines  typically  emphasize  the  importance  of their central theories \cite{Ralph:2014}. Software engineering (SE), in contrast, has no  widely  accepted  central  or  general  theories  \cite{pontus:2012}. SE  would  benefit  from  a  general  theory  in  numerous  ways \cite{Ralph:2014}: 1)  general  theories facilitate  developing  a  cumulative,  communicable  body  of  knowledge by preserving existing results and implicitly coordinating future studies; 2)  general  theories  improve  resistance  to  fads  by  stabilizing  the discipline; 3) general theory helps root out dubious and pseudoscientific concepts. 

Some research paths can be visualized regarding a consolidation of a Method Theory for MBSE:

\begin{itemize}
	\item \textbf{A consolidation of standard techniques for metamodeling}. Metamodeling is a core recurrent activity in MBSE. However, there is no consensus on the best techniques to perform it since several authors have proposed their own guidelines to construct metamodels \cite{jin:2001, essence:2001, Hug2009, deLara:2010, Li:2010, Cicchetti466155, Bak:2013, BergmayrW13, ArianNik2014, ClarkSW15, Sen2015, Piho2014457}. An investigation on this topic can foster methods consolidation for MBSE since its methods intensively depends on such activity;
	\item \textbf{A unified recommendation on how to transform models}. Some paradigms also are known for model transformations \cite{lncs:2014}. 	Model-to-model (M2M) transformations usually use graph patterns \cite{LaraG05, gramot2005_tax, Schuerr2008, Greenyer2012}. 	Model-to-text (or Model-to-code (M2C))  category is distinguished  between  visitor-based  and  template-based approaches \cite{Czarnecki:2003}. In former, a very basic code generation approach consists in providing some visitor mechanism to traverse the  internal  representation  of  a  model  and  write  code  to  a  text  stream to generate code. In latter, a template of a program in the target technology is filled with code transformed from a source model. Further research is necessary to establish a unifying approach for model transformation, regardless of the approach used;
\end{itemize}

Other activities need to be better specified regarding MBSE, such as the application of design patterns for metamodeling \cite{Cho:2011}, and the association with Component-Based Software Engineering \cite{ortiz:2014}. However, we believe that consolidating metamodeling and transformation techniques could properly benefit several other associations of MBSE with other initiatives, since metamodeling and model transformation are the heart and soul of MBSE \cite{sendall:2003}. In fact, several methods for MBSE have been proposed in the literature \cite{Ghani:2014:HRE}. MDA itself can be considered an specific MBSE method. However, MDA for examples lacks from recommendations on how to engineer a transformer in such a way that its architecture supports changes in the transformation rules without impacts for the transformer's architecture \cite{earlyaspects:2011}. Other initiatives do exist. The diversity of such distinct initiatives is also a claiming for a unifying methods theory for MBSE.  

\section{Final Remarks}\label{sec:MBSEfinal}

In this paper, we reported preliminary insights on the investigation of methods for Model-Based Software Development (MBSE). We claim that MBSE does not hold a method theory and that, with the realization of this for MBSE, well-grounded results could be generalized to the dawn of a general theory of methods for software engineering. That theory would integrate a future \textit{everything theory for software engineering}. This is only an embryonic reflexion to instigate the software engineering community on the emergency of this topic. Further research needs to be performed to suitably ground a theory for methods in MBSE, and for software engineering as a whole.

\bibliographystyle{alpha}

\end{document}